# High Entropy Alloy Catalytic Action on MgH$_2$ Hydrogen Storage Materials

**S. K. Verma[1], S. S. Mishra[1,2], N. K. Mukhopadhyay[3], T. P. Yadav[1,4]***

[1]Department of Physics, Institute of Science, Banaras Hindu University, Varanasi-221005, India.
[2]Department of Physics, Jagran Collage of Arts, Science & commerce, Saket Nagar, Kanpur-208014, UP, India.
[3]Department of Metallurgical Engineering, Indian Institute of Technology (Banaras Hindu University), Varanasi-221005, India
[4]Department of Physics, Faculty of Science, University of Allahabad, Prayagraj-211002, India.

## Abstract:

Magnesium hydride (MgH$_2$) is the mostly used material for solid-state hydrogen storage. However, their slow kinetics and highly unfavorable thermodynamics make them unsuitable for the practical applications. The current study describes the unusual catalytic action of a new class of catalyst, a high-entropy alloy (HEA) of Al$_{20}$Cr$_{16}$Mn$_{16}$Fe$_{16}$Co$_{16}$Ni$_{16}$, on the de/re-hydrogenation properties of MgH$_2$. The onset desorption temperature of MgH$_2$ is reduced significantly from 376 °C (for pristine MgH$_2$) to 338 °C when it is catalyzed with a HEA-based catalyst. On the other hand, a fast de/re-hydrogenation kinetics of MgH$_2$ was observed during the addition of HEA-based catalyst. It absorbs ~ 6.1 wt% of hydrogen in just 2 minutes at a temperature of 300 °C under 10 atm hydrogen pressure and desorbs ~ 5.4 wt% within 40 minutes. At moderate temperatures and low pressure, the HEA-based catalyst reduced desorption temperatures and improved re-hydrogenation kinetics. Even after 25 cycles of de/re-hydrogenation, the storage capacity of MgH$_2$ catalyzed with the leached version of HEA degrades negligibly.

*Corresponding author Email: tpyadav@allduniv.ac.in

## Introduction

The designing of novel hydrogen storage materials and technologies is essential for on-board applications and is a challenging task in the development of hydrogen economy [1–5]. The huge hydrogen capacity of Mg-based materials (especially MgH$_2$) (theoretical hydrogen capacity of 7.6 wt%) [6], outstanding reversibility [7], natural abundance, low cost, and non-toxicity, which attracted intensive investigation in the last several decades, make them one of the most promising choices for solid hydrogen storage[6,8–11]. However, magnesium has a very strong affinity for hydrogen and it forms a very stable Mg-H bond, which forms magnesium hydride (MgH$_2$) and needs high thermal energy in the form of enthalpy (theoretically ~76 kJ/mol-H$_2$) to decompose into Mg [8,12]. Due to the high desorption enthalpy, desorption requires an extremely high temperature (~ 400 °C), which is obviously undesirable in applications for proton exchange membrane fuel cells (PEMFCs)[12]. Additionally, the de/re-hydrogenation's kinetics is slow [13,14].

Over the past few decades, significant progress has been achieved to improve the de/re-hydrogenation properties, particularly the hydrogen release kinetics due to the enormous efforts put into catalysis, alloying, and nano-structuring [15–22]. The use of transition metals and their composites as catalyst, such as titanium, titanium hydride ($TiH_2$), niobium, vanadium, manganese, iron, cobalt, nickel, copper, chromium, palladium, and Al-Cu-Fe quasicrystalline alloys have been the subject of the most notable studies [6,10,16,23–28]. These metals and some of their compounds have been shown to exhibit significant improvement in the hydrogen sorption kinetics of $MgH_2$. Pandey et al.[26] have synthesized $MgH_2$ catalyzed by various versions of Al-Cu-Fe quasicrystalline materials. They have reported the improved de/re-hydrogenation kinetics that absorbed 6.00 wt% in 30 seconds at 250 °C under 20 atm $H_2$ pressure and released the 6.30 wt% hydrogen within 150 seconds at 320 °C at 1 atm $H_2$ pressure. Improved kinetics and a decrease in desorption activation energy were also noted by Ouyang et al. [29], but they were unable to detect any changes in enthalpy, even after numerous cycles. Our previous study [23], have reported that catalyzed $MgH_2$ shows the fast de/re-hydrogenation kinetics as well as lowering of thermodynamic barrier by ~ 6 kJ/mol $H_2$ (change in enthalpy) than that of the pristine $MgH_2$. However, in a recent study [10], we have reported the enhanced de/re-hydrogenation kinetics without significant changes in thermodynamic barriers. Moreover, it should be observed that the desorption temperature is still very high and the catalytic impact of the composite is not significant enough. This is true because the dehydrogenation process involves a gas-solid reaction, and the gas-solid interface has a significant impact on the reaction rate [12]. The $H_2$ molecule recombination on the surface of $MgH_2$ is therefore the most important problem during the dehydrogenation process of $MgH_2$.

On the other hand, researches have typically used the method of modifying the reaction path through the production of composites in the dehydrogenation process to improve thermodynamics, which entails lowering the reaction enthalpy. Both the MgAl alloys [30] and the solid solution of $Mg_{0.95}In_{0.05}$ [31] have produced some excellent results. $Mg_2In_{0.1}Ni$ solid solution phase has demonstrated the impact of modifying the reaction path in affecting the thermodynamics of de/re-hydrogenation by showing a considerable drop in enthalpy [31]. In comparison with $MgH_2$ catalyzed with metal catalysts, Shahi et al.[32] have demonstrated that the combined effect of various transition metals, such as Ti, Fe, and Ni, on the de/re-hydrogenation characteristics of $MgH_2$ (i.e., $MgH_2$-$Ti_5Fe_5Ni_5$), which offers superior hydrogen storage properties. They have demonstrated that the $MgH_2$-$Ti_5Fe_5Ni_5$ absorbed 5.30 wt% of hydrogen within 15 minutes at temperature of 270 °C under 12 atm $H_2$ pressure.

Nevertheless, maintaining ~ 6.00 wt% of storage capacity along with reversibility for $MgH_2$ is still an open problem that needs to be solved despite the aforementioned and other investigations. Furthermore, it is evident from the various studies mentioned above that metals and combinations of metals both functions well as catalysts for $MgH_2$, but a more effective strategy is still required to make $MgH_2$ a practical hydrogen storage material. Contrary to conventional alloys, which typically only contain one or two base elements, high-entropy alloys (HEAs) contain numerous principal elements, and the range of possible HEA compositions is far wider than that of conventional alloys [33]. HEAs are multicomponent alloys, the lattice strain in the alloy makes it favorable to absorb hydrogen in both tetrahedral and octahedral interstitial sites, and thus HEAs have the potential to be used as hydrogen storage materials in the future [34,35].

In view of the above, the present study has explored to develop a new class of catalyst that can improve the de/re-hydrogenation kinetics of MgH$_2$. Here, Al$_{20}$Cr$_{16}$Mn$_{16}$Fe$_{16}$Co$_{16}$Ni$_{16}$ HEA and its modified (leached) version (modification or leaching in this context refers to the partial removal of Al atoms on the surface of HEA via chemical processing by using NaOH as leaching agent) were utilized as catalyst for MgH$_2$. However, HEAs itself are in great discussion nowadays for hydrogen storage application but their poor hydrogen storage capacity needs further improvements [36–38]. It is well understood that the catalytic activity of any catalyst is determined by its structure, particle size, and electronic structure. A new kind of catalyst for hydrogen sorption in MgH$_2$ is thus anticipated to be formed by HEA materials, especially those that incorporate transition metals. As far as we are aware and to the best of knowledge, there are only few recent studies [39–41] where HEAs are used as catalyst for MgH$_2$ system. Moreover, there is no reported studies on using this kind of modified version of HEA catalyst to enhance MgH$_2$'s de/re-hydrogenation capabilities.

## Experimental section
**Synthesis protocol of as-cast Al$_{20}$Cr$_{16}$Mn$_{16}$Fe$_{16}$Co$_{16}$Ni$_{16}$ HEA and its leached version:**

The high-entropy alloy with nominal composition Al$_{20}$Cr$_{16}$Mn$_{16}$Fe$_{16}$Co$_{16}$Ni$_{16}$ (in at%) was synthesized using vacuum arc melting in inert argon atmosphere. The high purity constituents namely Al (purity 99.9%, Alfa-Aesar), Cr (purity 99.9%, Merck), Mn (purity 99.95%, Alfa-Aesar), Fe (purity 99.5%, Riedel de-haen) Co (purity 99.9%, Sigma Aldrich), and Ni (purity 99.99%, Sigma Aldrich) were used for the synthesis of HEA. The sample was re-melted five times, each time by flipping the button, to ensure homogeneity. The as-cast button ingot encapsulated within quartz tube in an inert atmosphere, was homogenized at 600 °C for 10 h and then quenched to cold water. The homogenized sample was ball milled for 24 h in planetary ball mill with 30:1 ball to powder ratio. In order to prepare the catalyst, the ball milled powder was leached with 2M NaOH aqueous solution. The sample was exposed to leaching for 2h in open atmosphere. The leached sample was washed with distilled water several times to remove the NaOH. The leached and washed HEA (LHEA) sample was dried at 50 °C under vacuum and then used as catalyst for MgH$_2$ system for further investigations.

**Synthesis route of catalyzed version of MgH$_2$:**

The MgH$_2$ (purity 99%) was obtained from Fujifilm (Japan) for the present study. To synthesize MgH$_2$ catalyzed by LHEA, LHEA and MgH$_2$ were mechanically milled at 200 rpm for 24 hours in a planetary ball-miller (Retsch PM 400; ball to powder ratio, 50:1). Ball-milling was carried out at a hydrogen pressure of 5 atm. We have prepared a set of three different concentrations of catalyst (LHEA) (5 wt%, 7 wt%, and 10 wt%) for catalyzing MgH$_2$. In terms of desorption temperature and hydrogen storage capacity (given in Fig. S2 in supplementary information file), it was found that 7 wt% of catalyst (LHEA) is optimum for hydrogen sorption in Mg/MgH$_2$. So, 7 wt% of the MgH$_2$ was assumed to be the optimum catalyst's concentration for the present study. All samples were handled inside a N$_2$-filled glove box (MBRAUM MB10 compact). For a comparative study of de/re-hydrogenation kinetics of MgH$_2$-LHEA, we also have investigated the responses of the pristineMgH$_2$.

**Characterization tools:** The prepared samples were structurally characterized using PANalytical Empyrean x-ray diffraction (XRD) system fitted with an area detector (256x256 pixels) with a CuKα line (λ = 1.5415 Å) operated at 40 kV, 40 mA. The Tecnai-20 $G^2$ transmission electron microscope (TEM) with 200 kV of accelerating voltage was used to record the sample's microstructures and selected area electron diffraction pattern (SAED). The surface morphology was examined by using a scanning electron microscope (SEM) (Model: FEI Quanta 200) with an operating voltage of 25 kV (vacuum of $10^{-5}$ torr), and the elemental composition of the prepared materials were analyzed by energy dispersive X-ray analysis (EDX) equipped with SEM. The X-ray photoelectron spectroscopy (XPS) analysis of the samples were performed using the Physical Instruments (PHI) 5000 Versa Probe-III spectrometer with AlKα radiation (1486.6 eV) under an extremely high vacuum ($10^{-12}$ torr). The dehydrogenation characteristics of the as-prepared samples were evaluated using temperature programmed desorption (TPD) at a heating rate of 5 °C/min. An automated four-channel volumetric sieverts type apparatus was used to perform all of the de/re-hydrogenation measurements (Advanced Materials Corporation Pittsburgh, USA). The samples were heated from room temperature to approximately 450 °C with different heating rates of 10 °C/min, 12 °C/min, 15 °C/min, and 18 °C/min under flowing nitrogen (20 ml/min), and then thermally analyzed using a differential scanning calorimeter (DSC 800, Perkin Elmer). The de/re-hydrogenation kinetics and thermodynamics of $MgH_2$ catalyzed by HEA, leached high-entropy alloy (LHEA) and a feasible corresponding mechanism of catalysis have been put forward. The following are the abbreviations used in the present investigation as given in Table 1.

**Table: 1.** *Abbreviation used in the present investigation.*

| S. No. | Composite | Abbreviation |
|---|---|---|
| (a) | As-cast $Al_{20}Cr_{16}Mn_{16}Fe_{16}Co_{16}Ni_{16}$ high-entropy alloy | HEA |
| (b) | Leached $Al_{20}Cr_{16}Mn_{16}Fe_{16}Co_{16}Ni_{16}$ high-entropy alloy | LHEA |
| (c) | $MgH_2$ catalyzed by as-cast 7wt% $Al_{20}Cr_{16}Mn_{16}Fe_{16}Co_{16}Ni_{16}$ high-entropy alloy | $MgH_2$-HEA |
| (d) | $MgH_2$ catalyzed by 7wt% leached $Al_{20}Cr_{16}Mn_{16}Fe_{16}Co_{16}Ni_{16}$ high-entropy alloy | $MgH_2$-LHEA |

## Results and discussion

### Structural, microstructural, and surface morphology analysis

The XRD patterns of as-cast and leached versions of HEA material describe the structural information of the catalyst as shown in Fig. 1. As-cast $Al_{20}Cr_{16}Mn_{16}Fe_{16}Co_{16}Ni_{16}$ HEA exhibits face centered cubic (FCC, a = 3.701 Å) structure, identified by the presence of representative prominent diffraction peak at 2θ≈44.06º as (111) (shown in Fig. 1(a)). Fig. 1(b) represents the XRD pattern of pristine $MgH_2$, where all the peaks matches well with the tetragonal lattice structure of space group P42/mnm(136) and lattice parameters a=b=4.516Å, c=3.020Å (joint committee on powder diffraction standards (JCPDS) no. 740934). The single-phase FCC lattice structure of leached version of HEA can be identified in the Fig. 1(c). The XRD pattern of $MgH_2$-HEA is shown in Fig. 1(d), where all the diffraction peaks matches well with the tetragonal phase of $MgH_2$ along with the two diffraction peaks of HEA at 2θ ≈ 26.50º and 44.06º. From Fig. 1(e), the usual peaks of $MgH_2$ catalyzed by LHEA can be seen besides the tetragonal phase of $MgH_2$, the signature of LHEA is observed in the form of diffraction peak at 2θ =

44.06°. The crystallite size of prepared samples have been estimated by using Debye Scherer's formula [42]. The average crystallite size of as-cast HEA, LHEA, pristine MgH$_2$, and milled MgH$_2$ are determined to be ~ 49 nm, ~7 nm, ~88 nm, and ~10 nm respectively.

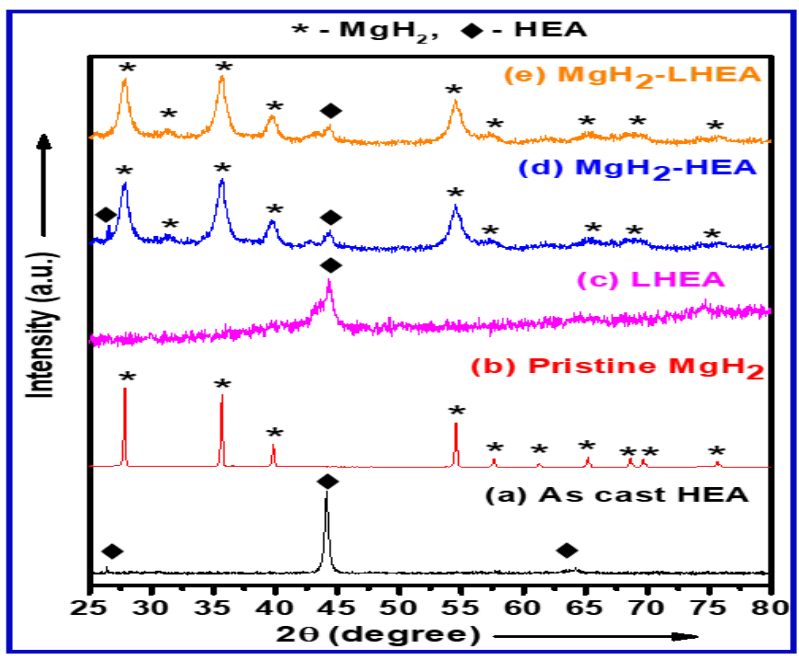

**Fig. 1:** *XRD patterns for (a) as-cast HEA, (b) pristine MgH$_2$, (c) LHEA, (d) MgH$_2$-7wt%HEA, and (e) MgH$_2$-7wt%LHEA.*

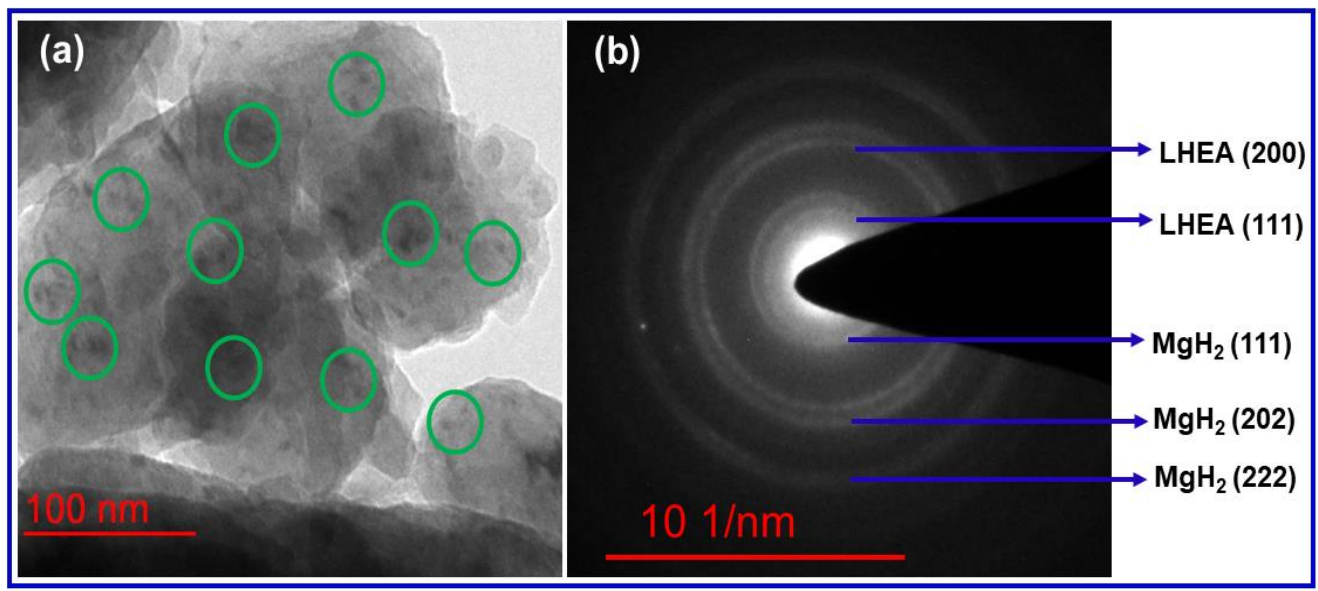

**Fig. 2:** *(a) TEM micrograph of MgH$_2$-7wt%LHEA; catalyst particles encircled by green circles and (b) corresponding SAED pattern.*

To probe the microstructural details of the sample, TEM has been utilized. The representative TEM micrograph and SAED pattern of MgH$_2$-LHEA sample has been shown in Fig. 2. TEM image of MgH$_2$-LHEA sample has been shown in the Fig. 2(a), reveals the uniform distribution of LHEA catalyst particles (marked by green circles) on the MgH$_2$ matrix, where the estimated catalyst particle size observed in the range of 18-23 nm. The presence of catalyst particles in the MgH$_2$ matrix can also be confirmed by the SAED pattern of MgH$_2$-LHEA (shown in Fig. 2(b)), where the diffuse SAED ring pattern indicating the existence of very short-range crystallinity in the materials was observed in TEM micrograph.

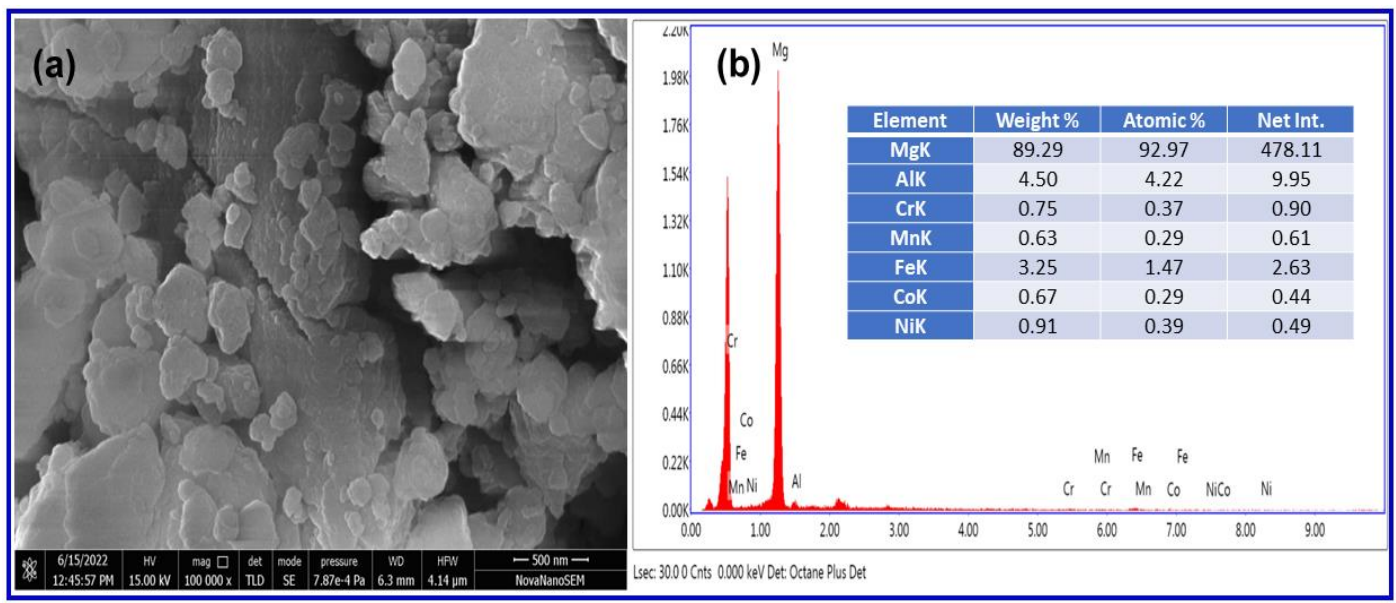

**Fig. 3:** *(a) SEM micrograph of MgH$_2$-7wt%LHEA and (b) EDX spectra with elemental percentage composition analysis.*

The surface morphology and elemental composition of the MgH$_2$-LHEA sample was investigated by SEM and EDX analysis respectively (shown in Fig. 3). Fig. 3(a) shows the surface morphology of the MgH$_2$-LHEA sample, where the cracks on the material surface can be attributed to the expansion of lattice due to presence of H$_2$ absorption during milling. The presence of LHEA catalyst in the MgH$_2$ may also be verified by the EDX spectra and corresponding elemental composition in the material (shown in Fig. 3(b)).

**XPS spectroscopy analysis**

The results of the XPS analysis can be used to assess the sample's elemental makeup, chemical states, and electronic states. Generally, the 2p core level peaks spectrum are recorded for alloys containing transition metals [43]. The samples of hydrogenated MgH$_2$-LHEA and dehydrogenated MgH$_2$-LHEA have been characterized by XPS spectroscopy, and peaks have been deconvoluted, in order to obtain information about the electronic environment of the catalyst employed for this experiment. Fig. 4 and 5, displays the deconvoluted peaks of XPS spectra for hydrogenated MgH$_2$-LHEA and dehydrogenated MgH$_2$-LHEA, respectively. The Al 2p core level spectrum of re/de-hydrogenated MgH$_2$-LHEA samples is fitted with a single peak of 2p$_{3/2}$ spectral line at ~ 72.2 eV corresponds to +3 oxidation state which reveals there is no change in electronic state of Al during re/de-hydrogenation process [44].

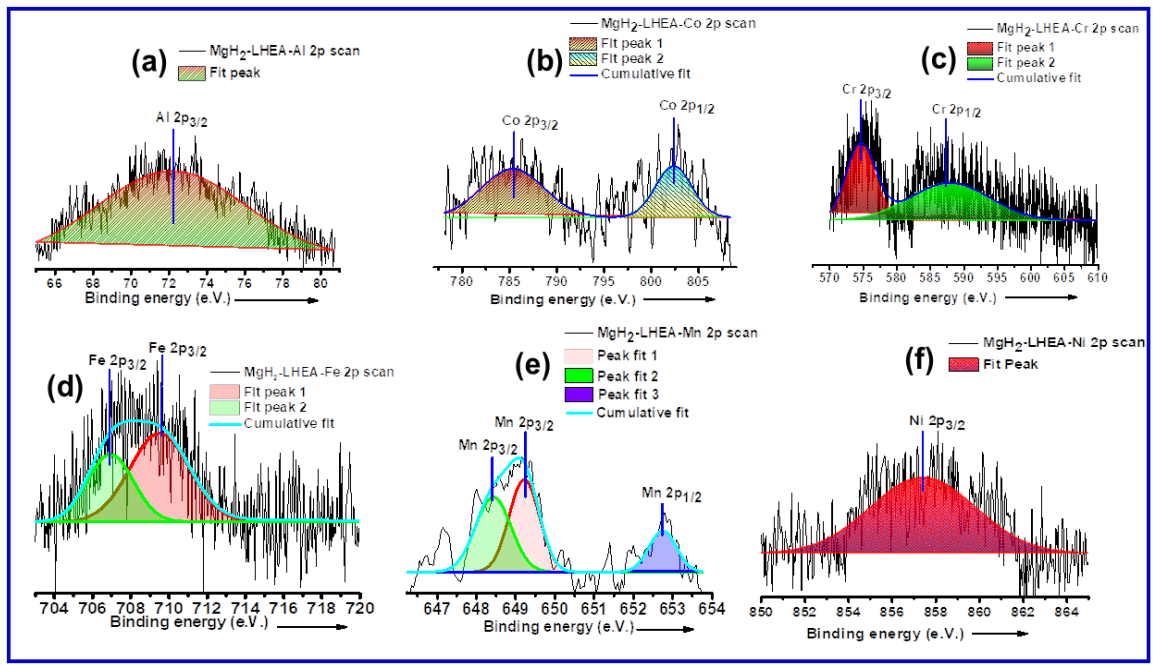

**Fig. 4:** *XPS spectra plot for 2p regions of catalyst elements of rehydrogenated MgH$_2$-7wt%LHEA.*

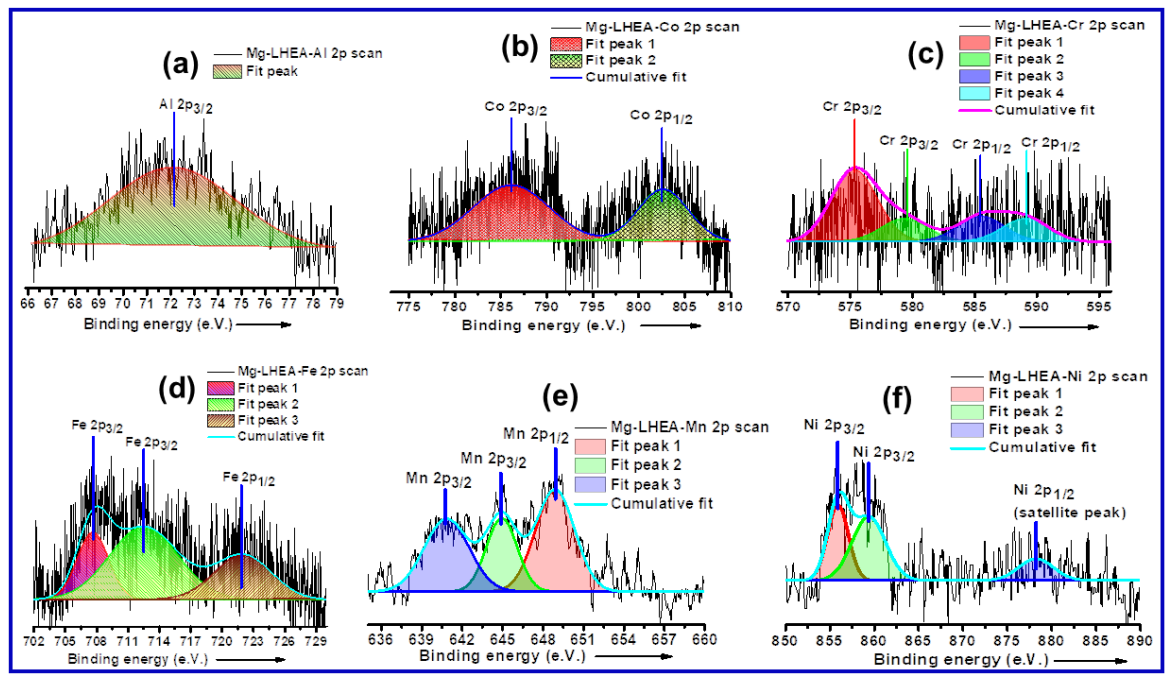

**Fig. 5:** *XPS spectra plot for 2p region of catalyst elements of dehydrogenated MgH$_2$-7wt%LHEA (that is Mg-LHEA).*

In the case of Co, the 2p spectrum is fitted with two peaks at ~ 783.1 eV ($2p_{3/2}$) and ~ 802.4 eV ($2p_{1/2}$) corresponds to +2 oxidation state [45,46] for rehydrogenated sample while in dehydrogenated sample Co 2p spectrum is fitted in two peaks at ~ 782.4 eV ($2p_{3/2}$) and ~ 802.6 eV ($2p_{1/2}$) corresponds to +3 and +2 oxidation states [45,46] respectively. The +3-oxidation state [47,48] is observed for Cr 2p spectrum (574.6 eV ($2p_{3/2}$), 587.3 eV ($2p_{1/2}$)) for hydrogenated $MgH_2$-LHEA sample whereas variable oxidation states +4 (575.4 eV ($2p_{3/2}$)) [49], +3 (579.6 eV ($2p_{3/2}$)) [50], +2 (585.5 eV ($2p_{1/2}$)) [51], +6 (589.2 eV ($2p_{1/2}$)) [48] of Cr 2p spectrum for dehydrogenated sample were observed. The Fe 2p spectrum for hydrogenated sample is fitted with two peaks of 706.4 eV and 710.6 eV ($2p_{3/2}$) corresponds to +2 oxidation states [52] while Fe 2p spectrum for dehydrogenated sample were fitted with three different peaks 707.6 eV ($2p_{3/2}$), 712.5 ($2p_{3/2}$), 721.7 ($2p_{1/2}$) of oxidation states +2, +3, +3 [52–54] respectively. On the other hand, Mn 2p at 648.5 eV ($2p_{3/2}$), 649.3 eV ($2p_{3/2}$), 652.7 eV shows the +7, 0, +2-oxidation states, respectively [55] in the hydrogenated sample while +2, +4, 0 oxidation states corresponds to peak positions 640.8 eV, 644.9 eV, 648.9 eV [55–57] in the dehydrogenated sample. The +2-oxidation state [58] of Ni (857.4 eV ($2p_{3/2}$)) is shown in hydrogenated sample whereas +3 (855.9 eV ($2p_{3/2}$)), +2 (859.3 eV ($2p_{3/2}$)), and +2 (878.6 eV) oxidation states [58–60] of Ni were observed in the case of dehydrogenated sample. The spectral lines and corresponding peak positions with electronic states of the re/de-hydrogenated $MgH_2$-LHEA can be seen in the table ST1 (shown in supplementary information). On the basis of XPS analysis (Fig. 4, 5, and table ST1), it is clear that catalyst (LHEA) changes its electronic states from rehydrogenation to dehydrogenation. Due to change in electronic states of catalyst during re/de-hydrogenation it provides a multivalent environment to $MgH_2$. This multivalent environment is helpful to destabilize the Mg-H bond resultant a fast re/dehydrogenation kinetics was obtained.

**De/re-hydrogenation kinetics study:**

The TPD was subsequently performed for the as-prepared samples with a constant heating rate of 5 °C/min, as shown in Fig. 6.

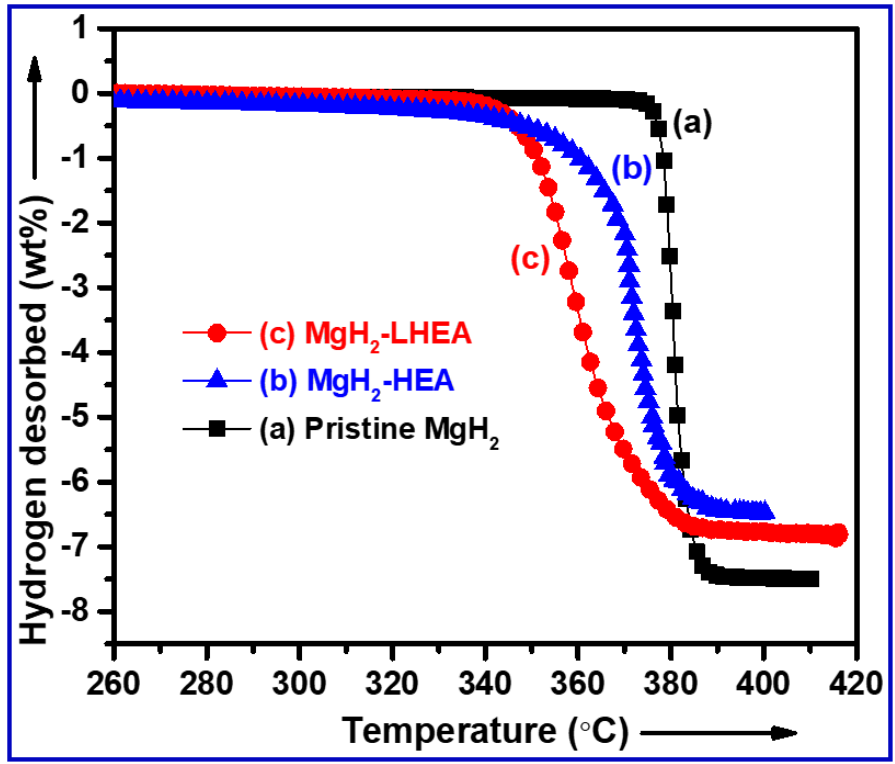

Fig. 6: *TPD curves for (a) pristine $MgH_2$, (b) $MgH_2$-7wt%HEA, and (c) $MgH_2$-7wt%LHEA.*

The MgH$_2$-LHEA starts hydrogen desorption at temperature of 338°C, and the hydrogen desorption is completed at 387°C with a desorption of ~6.8 wt% hydrogen (Fig. 6(c)) whereas the dehydrogenation of MgH$_2$ catalyzed by as cast HEA (MgH$_2$-HEA) occurred at 345 °C with a hydrogen storage capacity of 6.4 wt% (as shown in Fig. 6(b)). The hydrogen storage characteristic of catalyzed samples is then compared using the TPD of pure MgH$_2$ (shown in Fig. 6(a)). The beginning desorption temperature of the pure MgH$_2$ is 376°C, and the total storage capacity released is around 7.4 wt%. Because of this, it is evident from the TPD investigation that MgH$_2$-LHEA has a lower desorption temperature of 38°C and 7 °C than that of the pristine MgH$_2$ and MgH$_2$-HEA samples, respectively.

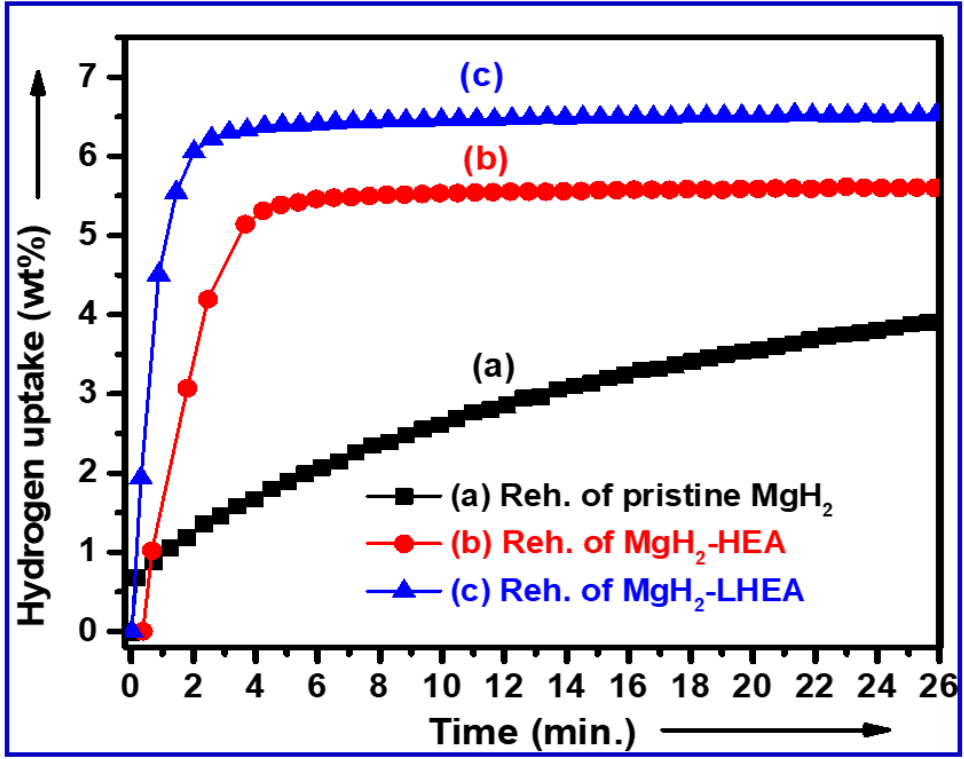

**Fig.7:** *Rehydrogenation curves for (a) pristine MgH$_2$ and (b) MgH$_2$-7wt%LHEA at a temperature of 300 ºC under 10 atm H$_2$ pressure.*

The hydrogen desorbed samples were then processed to check re/de-hydrogenation kinetics of catalyzed and uncatalyzed MgH$_2$. All the re-hydrogenation kinetics was conducted at 300 °C with hydrogen pressures of 10 atm. As can be seen from Fig. 7(a), at similar temperature and pressure conditions, the pristine MgH$_2$ absorbed ~ 1.3 wt% of hydrogen in two minutes while MgH$_2$-HEA absorbed ~3.57 wt% of hydrogen. On the other hand, 6.1 wt% of hydrogen was absorbed by MgH$_2$-LHEA sample in two minutes under the similar temperature and pressure conditions. Therefore, under the same temperature and pressure conditions, MgH$_2$-LHEA absorbs ~ 4.8 wt% and 2.53 wt% more hydrogen than pristine MgH$_2$ and MgH$_2$-HEA within 2 minutes. In comparison to pristine MgH$_2$, the data above demonstrate that MgH$_2$-LHEA exhibits the fast re-hydrogenation kinetics. The re-hydrogenated samples were then dehydrogenated at the temperature of 300 °C and 1 atm hydrogen pressure. In contrast to pristine MgH$_2$, and MgH$_2$-HEA, which releases ~ 0.1 wt%, 4.58 wt% of hydrogen within 40 minutes, and the MgH$_2$-LHEA sample releases ~ 5.4 wt% of hydrogen under the same time and similar temperature, and pressure conditions (shown in Fig. 8). From above mentioned de/re-hydrogenation kinetics study, it is clear that MgH$_2$ shows significantly faster hydrogen uptake and release in the presence of LHEA catalyst.

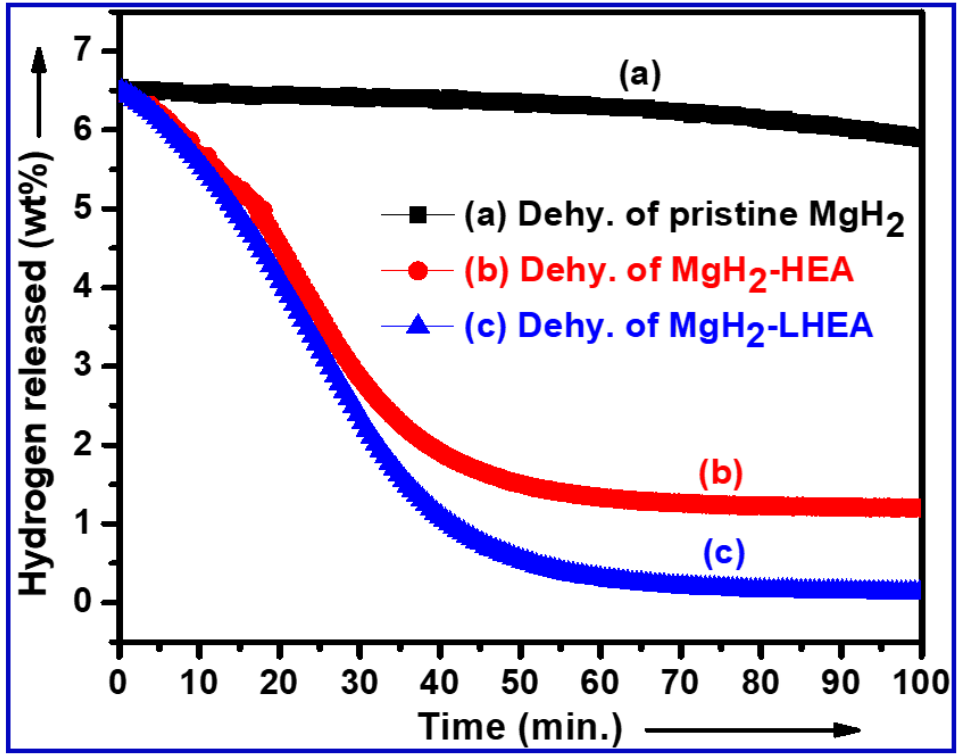

**Fig. 8:** *Dehydrogenation kinetics of (a) pristine MgH$_2$ and (b) MgH$_2$-7wt%LHEA at a temperature of 300 °C under 1 atm H$_2$ pressure.*

**Estimation of activation energy:**

The requirement of energy to overcome energy barrier for converting MgH$_2$ into Mg and vice-versa, (that is activation energy) were calculated using DSC analysis. Fig. 9 shows the DSC profiles of MgH$_2$-LHEA. As can be seen, the peak desorption temperature for MgH$_2$-LHEA was determined to be 406°C, whereas the onset hydrogen desorption temperature for MgH$_2$-LHEA was determined by TPD to be 338°C. Desorption temperature in the TPD (Fig. 6(b)) and DSC (Fig. 9(a)) curves differs because of several factors. One of the factors may be due to the fact that TPD was carried out in a vacuum, whereas the DSC was carried out in an environment of N$_2$. Using the Kissinger equation[61] (equation 1), we have performed DSC with a series of different sample heating rates (15, 18, 21, 24 °C/min) and plotted the Kissinger curve to determine the desorption activation energy.

$$\ln(\beta/T_p^2) = \left(-\frac{E_a}{RT_p}\right) + \ln(k_o) \ldots\ldots\ldots\ldots..(1);$$

where the symbols meanings are β-heating rate, Tp-peak desorption temperature, E$_a$-activation energy, R-gas constant (8.314 J/mol-K), and k$_o$-constant. The desorption activation energy is calculated using the slope of the ln(β/Tp$^2$) vs. 1000/Tp plot (Fig. 9(b)). According to calculations, the activation energy of MgH$_2$-LHEA is 121.21 kJ/mol (± 6.56 kJ/mol). According to this activation energy, it takes 121.21 kJ/mol of energy to break through the barrier that prevents MgH$_2$ from being converted into Mg in the presence of LHEA catalyst. The calculated activation energy for MgH$_2$-LHEA is much lower than the activation energy of pristine MgH$_2$.

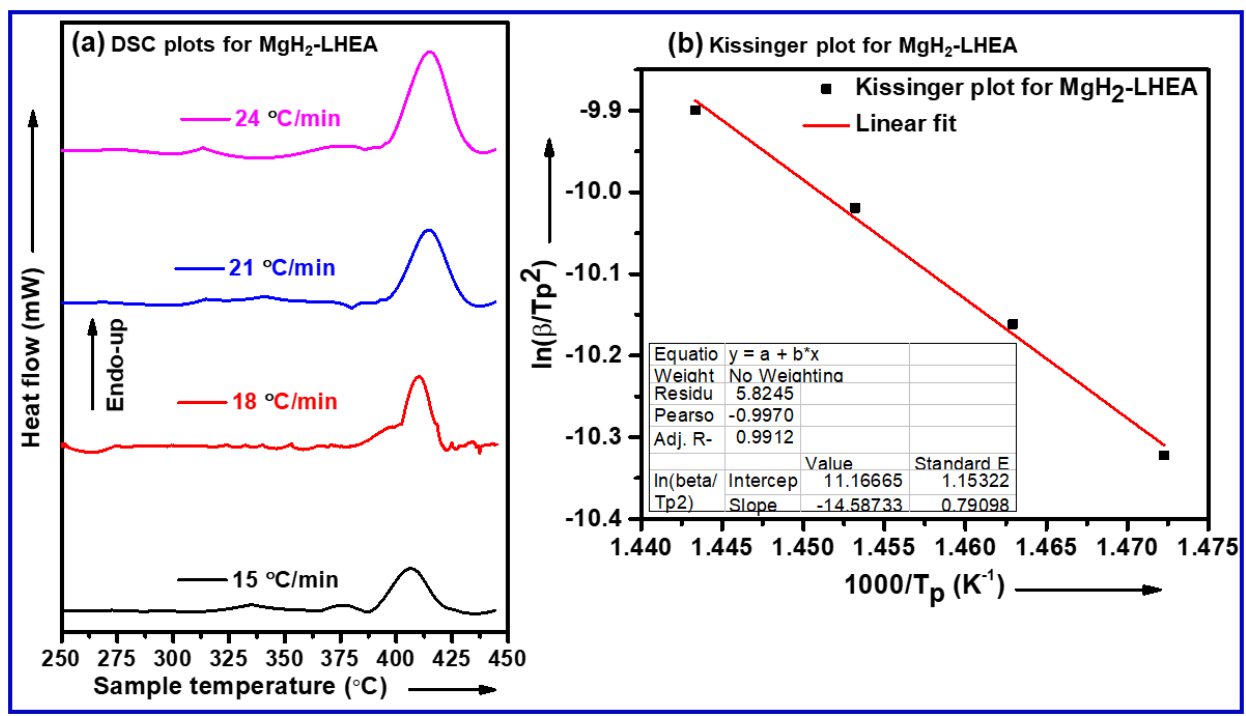

**Fig. 9:** *(a) DSC plots under various heating rates and (b) corresponding Kissinger plot for MgH$_2$-7wt%LHEA.*

**Study of thermodynamics: estimation of change in enthalpy (ΔH):**

After de/re-hydrogenation investigations, we next moved on to the thermodynamic analysis of catalyzed MgH$_2$ to compare the change in enthalpy (shown in Fig. 10). The change in enthalpy of MgH$_2$-LHEA was calculated using the PCI isotherms (Fig. 10(a)) and Van't Hoff plots (Fig. 10(b)). The estimated change in desorption enthalpy in the case of MgH$_2$ catalyzed by LHEA is 78.45 kJ/mol (±1.44 kJ/mol), which is nearly same as the theoretical hydrogen desorption enthalpy of pristine MgH$_2$. The data make it obvious that the presence of a catalyst does not result in any appreciable enthalpy change. Therefore, catalyst LHEA does not alter the thermodynamic parameter of MgH$_2$.

**Cyclic stability analysis:**

In light of the aforementioned studies, the LHEA catalyst (optimum catalyst) is crucial in enhancing the kinetics of MgH$_2$. In addition to kinetics and thermodynamics, the cyclic stability of the hydride material (MgH$_2$) is a crucial that makes it a respectable hydrogen storage material. The cyclic stability of the catalyzed MgH$_2$ sample must be examined as a result. To ensure the cyclic stability of catalyzed MgH$_2$, we carried out 25 cycles of dehydrogenation (under 1 atm hydrogen pressure at 300 °C) and re-hydrogenation (under 10atm hydrogen pressure at 300 °C). Fig. 11, depicts the cyclic stability curves for MgH$_2$-LHEA. It is evident from Fig. 11, that MgH$_2$-LHEA has a slight loss of storage capacity during re-hydrogenation, which is 0.07 wt% (from 6.57 wt% to 6.50wt%), and during dehydrogenation, which is 0.05 wt% (from 6.55 wt% to 6.50 wt%). With respect to cyclic stability, MgH$_2$-LHEA is significantly more stable than pristine MgH$_2$ at the same temperature and pressure.

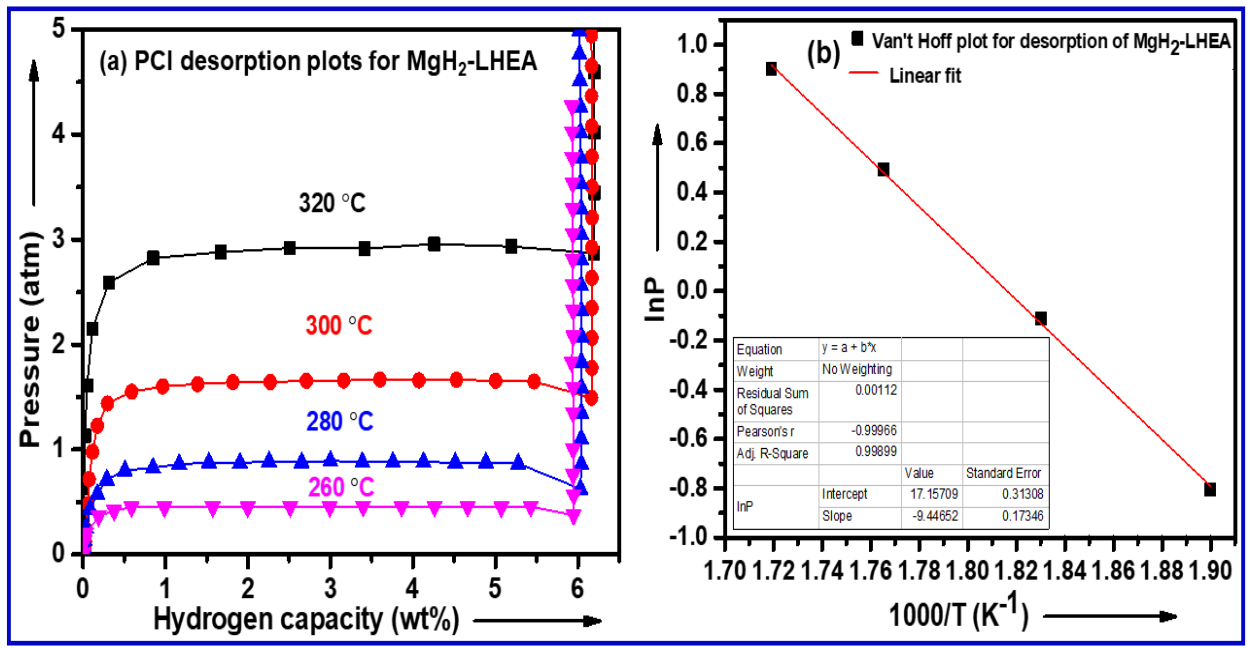

**Fig. 10:** *(a) PCI desorption plots at four different temperatures and (b) corresponding Van't Hoff plot for MgH$_2$-7wt%LHEA.*

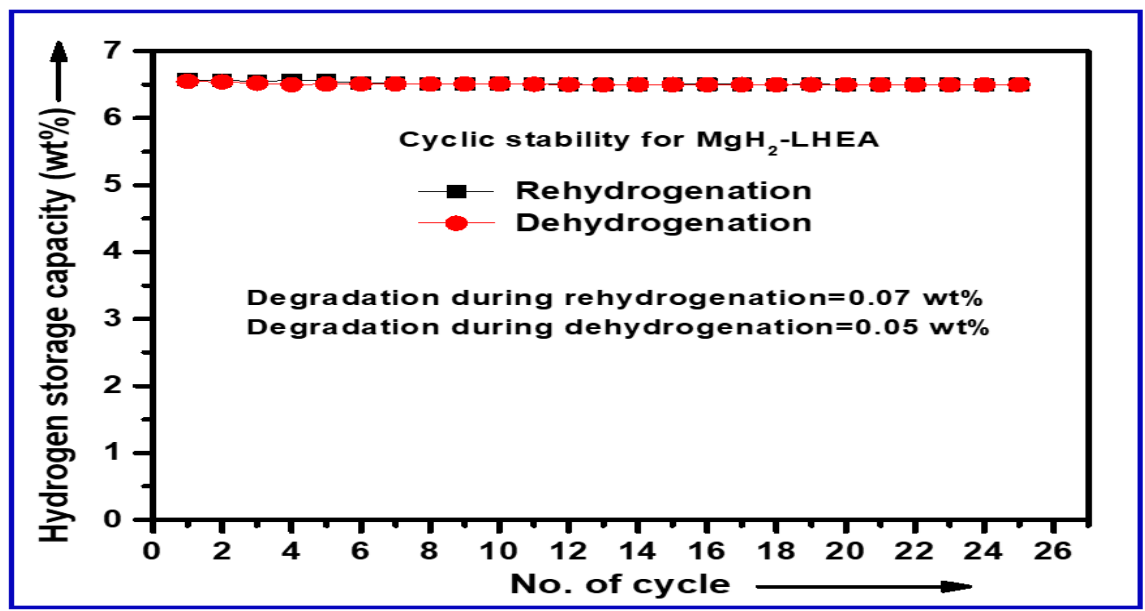

**Fig. 11:** *Cyclic stability curves for MgH$_2$-7wt%LHEA upto 25 cycles of re/de-hydrogenation. The rehydrogenation was performed at a temperature of 300 ºC under 10 atm H$_2$ pressure while dehydrogenation at 300 ºC under 1 atm H$_2$ pressure.*

To justify the importance of the present study, we have compared its results with some previous reported studies where the various alloys are used as catalyst to catalyze MgH$_2$ for hydrogen storage application. The comparison of results is tabulated in Table 2.

**Table: 2.** *Comparison of hydrogen storage properties for MgH$_2$ catalyzed by various transition metal-based alloys.*

| S. No. | Material composition | Method of catalyst processing | On set dehydrogenation temperature (°C) | Hydrogen storage capacity (wt%) | Dehydrogenation activation energy (kJ/mol-H$_2$) | Cyclic stability with capacity degradation | Hydrogen release kinetics | Ref. |
|---|---|---|---|---|---|---|---|---|
| 1. | MgH$_2$-9%FeCoNiCrMo | As cast | 200 | 6.7 | 84.47 | 20 cycles (0.28 wt%) | Desorbed 5.2 wt% (17 min, 325 °C) | [62] |
| 2. | MgH$_2$-5%FeCoNiCrMn | As cast | 275 | 6.1 | 90.20 | 50 cycles (0.08 wt%) | Desorbed 4.5 wt% (40 min, 240 °C) | [40] |
| 3. | MgH$_2$-10%CrFeCoNi | As cast | 232 | 6.5 | 133.06 | 20 cycles (0.20 wt%) | Desorbed 5.6 wt% (10 min, 300 °C) | [41] |
| 4. | MgH$_2$-10%TiVNbZrFe | As cast | 200 | 5.78 | 63.03 | 100 cycles (0.32 wt% in 15 cycles) | Desorbed 5.25 wt% (30 min, 270 °C) | [39] |
| 5. | MgH$_2$-50%NiMnAlCoFe | As cast | 180 | 2.0 | 131.34 | -- | -- | [63] |
| 6. | MgH$_2$-10%TiCuNiZr | As cast | 250 | 5.5 | 104.70 | -- | Desorbed 4.9 wt% (5 min, 300 °C) | [64] |
| 7. | MgH$_2$-20%TiCrMnV | As cast | 294 | 6.0 | 71.20 | -- | -- | [65] |
| 8. | MgH$_2$-7%HEA | As cast | 345 | 6.4 | -- | -- | Desorbed 4.58 wt% (40 min., 300 °C) | Present study |
| 9. | MgH$_2$-7%LHEA | Leached version of HEA | 338 | 6.8 | 121.21 | 25 cycles (0.05 wt%) | Desorbed 5.4 wt% (40 min, 300 °C) | Present study |

**Tentative mechanism of catalytic action of LHEA on MgH$_2$:**

The tentative mechanism of LHEA for de/re-hydrogenation in MgH$_2$/Mg can be understood with the help of the XRD, TEM, SEM-EDX, and XPS analysis. As can be seen by XRD patterns (Fig. 1) and Debye Scherer's analysis (in Structural, microstructural, and surface morphology analysis section), it is clear that leaching of HEA followed by milling provides material in the form of small size catalyst with average crystallite size is 7 nm whereas the average crystallite size of as cast HEA was ~49 nm. These nano-catalyst (LHEA) particles distributed homogeneously within the MgH$_2$ matrix by ball-milling, which creates various hydrogen diffusion channels. These diffusion channels accelerate the movement of atomic hydrogen from/in MgH$_2$/Mg which results in reducing the kinetic energy of de/re-hydrogenation [66]. Figure S1 (in ESI file) depicts the typical XRD pattern of MgH$_2$-LHEA after the first and twenty-fifth cycles of dehydrogenation. It should be noticed that even after 25 cycles, there has been no appreciable change in the full width at half maximum (FWHM) of the XRD peaks of Mg and LHEA. This implies that no grain growth is taking place. As a result, there is no LHEA nanoparticles aggregation. Therefore, the cyclic stability of MgH$_2$-LHEA will be preserved that can be seen in 'cyclic stability' section where 0.05 wt% of storage capacity was degraded up to 25$^{th}$ cycles. Due to partial leaching of Al atoms on the surface of HEA (Al$_{20}$Cr$_{16}$Mn$_{16}$Fe$_{16}$Co$_{16}$Ni$_{16}$), the concentration of Fe atoms on the surface is increased that can be seen in the SEM-EDX spectra and corresponding elemental composition (given in Fig. 3(b)). As it is well documented in a previous study [67] that Fe works as bridge for electron transfer and the property of electron attraction strength make it the prominent candidate for destabilizing the Mg-H bond. It is also previous reported that stand-alone Fe, Co, Ni, Cr, and Mn have good catalytic effects on the MgH$_2$ hydrogen storage system [66,68–71]. When the atomic sizes of the constituent elements are similar, HEA is more likely to produce a stable solid solution structure. The "cocktail effect" causes the catalytic effects of Cr, Mn, Fe, Co, and Ni to synergistically work together to achieve enhanced catalytic effect by producing high-entropy alloys. Additionally, the surface is very heterogeneous due to the multi-element arrangement, which is advantageous for offering a variety of active sites for hydrogen diffusion during the dehydrogenation and rehydrogenation of MgH$_2$. The de/re-hydrogenation from/in MgH$_2$/Mg depends critically on the stability of the Mg-H bond; the more quickly it destabilizes, the more quickly the de/re-hydrogenation proceeds. On the other hand, partial leaching of Al atoms on the surface of HEA makes the material (catalyst) environment in multi-electronic states that can be helpful to destabilize the Mg-H bond. Now the question arises: how the partial leaching of Al-atoms on the surface of HEA makes the catalyst (LHEA) more multivalent environment? To answer this, we have characterized the rehydrogenated and dehydrogenated MgH$_2$-LHEA samples using XPS. In order to determine the oxidation states of Al, Co, Cr, Fe, Mn, and Ni, in the MgH$_2$-LHEA and Mg-LHEA, the XPS analysis of MgH$_2$-LHEA and Mg-LHEA was performed. The deconvoluted XPS spectra of MgH$_2$-LHEA and Mg-LHEA are shown in Fig. 4 and 5, respectively and results tabulated in Table ST1. The presence of spectral lines 2p$_{3/2}$ of Al in MgH$_2$-LHEA and Mg-LHEA corresponds to the Al +3 electronic state (shown in Figs. 4(a), 5(a) and Table ST1). The electronic states of Co 2p$_{3/2}$, Co 2p$_{1/2}$ present in the MgH$_2$-LHEA sample correspond to Co +2, whereas the electronic state of Co in Mg-LHEA corresponds to the presence of Co +3 and Co +2 (shown in Figs. 4(b), 5(b) and Table ST1). The Cr 2p$_{3/2}$ and Cr 2p$_{1/2}$ spectral lines in MgH$_2$-LHEA indicates the Cr +3 oxidation sate while Cr +2, Cr +3, Cr +4, and Cr +6 oxidation states were observed in Mg-LHEA sample (shown in Figs. 4(c), 5(c) and Table ST1). The Fe 2p$_{3/2}$ in the MgH$_2$-LHEA sample shows Fe +2 electronic state, while in Mg-LHEA, Fe possesses +2 and +3 electronic states (shown in Figs.

4(d), 5(d) and Table ST1). The Mn $2p_{3/2}$ spectral line of $MgH_2$-LHEA shows +7, 0, +2 oxidation states while Mn +4, +2, 0 oxidation states were observed in Mg-LHEA sample (shown in Figs. 4(e), 5(e) and Table ST1). The Ni $2p_{3/2}$ spectral line of $MgH_2$-LHEA corresponds to +2 oxidation state while +2 and +3 oxidation states of Ni was observed in Mg-LHEA sample (shown in Figs. 4(f), 5(f) and Table ST1). On the basis of XPS analysis (Fig. 4, 5, and table ST1), it is clear that catalyst (LHEA) changes its electronic states during rehydrogenation to dehydrogenation process and more multi-electronic states of Co, Cr, Fe, Mn, and Ni could be observed in dehydrogenated sample. Due to change in multi-electronic states and higher electronic states of catalyst during re/de-hydrogenation it provides a multivalent environment to $MgH_2$. This multivalent environment is helpful to destabilize the Mg-H bond resultant a fast re/de-hydrogenation kinetics was obtained. According to earlier studies [12,72,73], a catalyst with a more multi-electronic state facilitates easier electron transfers (to and fro) and destabilizes the M-H bond more quickly than a catalyst with a less multi-electronic environment. Since the catalyst LHEA has more multi-electronic states of Co, Cr, Fe, Mn, and Ni during dehydrogenation than rehydrogenation of $MgH_2$. So, LHEA can destabilize the Mg-H bond of $MgH_2$ during dehydrogenation. Therefore, in the case of $MgH_2$-LHEA, de/re-hydrogenation kinetics get faster. The graphical understanding of the catalytic mechanism of LHEA over the $MgH_2$, we have given a schematic for catalytic mechanism of LHEA on $MgH_2$ in Fig. 12.

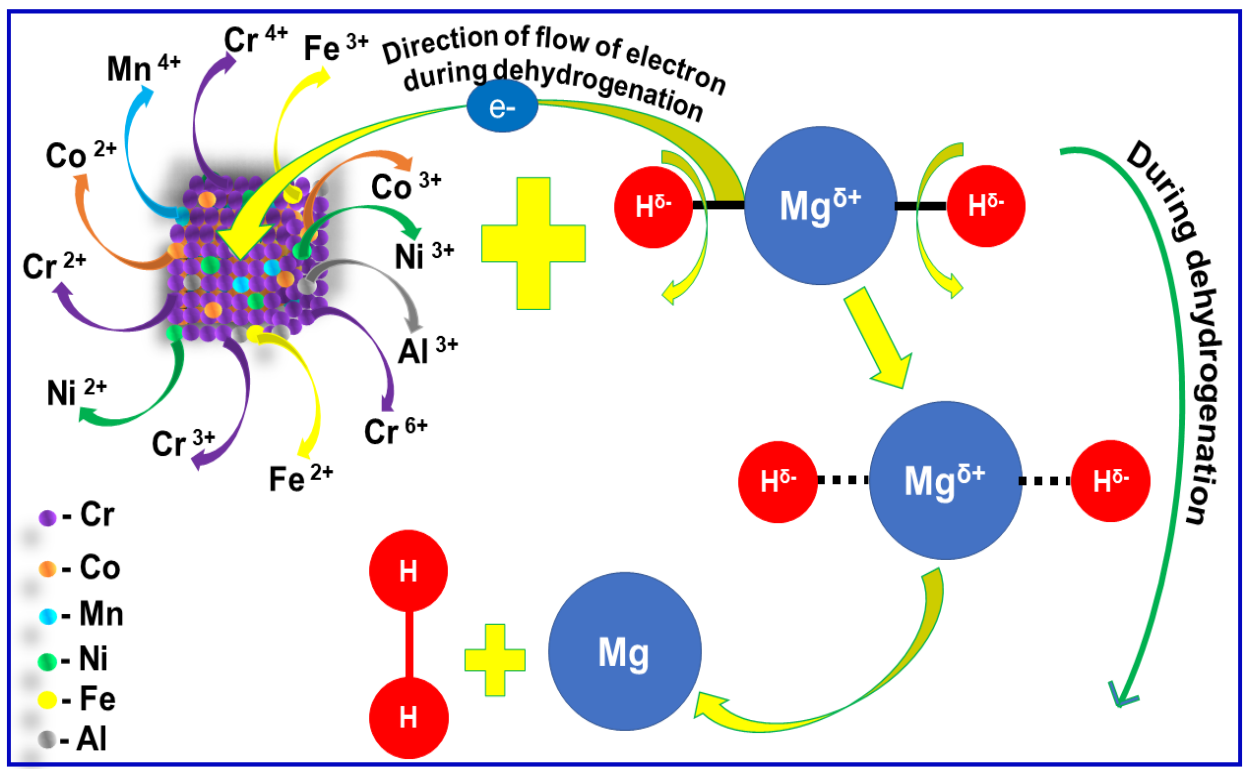

**Fig. 12:** *Schematic representation of catalytic mechanism of catalyst (LHEA) during dehydrogenation of $MgH_2$.*

## Conclusions

A new class of modified HEA catalyst is reported for the present investigation to catalyze the $MgH_2$ for hydrogen storage properties. According to the de/re-hydrogenation investigation, LHEA catalyst performed very well for $MgH_2$ over a time. The dehydrogenation temperature of $MgH_2$-LHEA is to be around 338 °C, which is 77 °C and 7 °C lower than that of the pristine $MgH_2$ and $MgH_2$-HEA, respectively. The $MgH_2$-LHEA sample absorbed the hydrogen ~ 6.1 wt% within 2 minutes and desorbed the ~5.4 wt% of hydrogen within 40 minutes at 300 °C. It also showed the strong cyclic stability up to 25 cycles with a minimum degradation of hydrogen storage capacity ~0.05 wt%. However, LHEA catalyst does not change significantly on the thermodynamic parameter ($\Delta H$) of $MgH_2$. In the presence of LHEA catalyst, it was observed that the hydrogen desorption activation energy barrier to converting $MgH_2$ into Mg is around 121.21 kJ/mol, which is significantly lower than that of the pure $MgH_2$. Studies conducted for this inquiry show that LHEA can be considered to be a key catalyst for $MgH_2$ that may open new direction for catalyzing various hydrides due to its multi-component behavior.

**Acknowledgements:**

This article is dedicated to distinguish researcher in the field of Hydrogen Energy, late Prof. O. N. Srivastava for his constructive suggestions for useful experiments, who departed this world on April 24, 2021. The authors are grateful for stimulating discussions with Professor M.A. Shaz. Author (S.K.V.) would like to thank the Council of Scientific and Industrial Research (CSIR) New Delhi, India for providing financial assistance as CSIR-Senior Research Fellowship (Award No. 09/013(0872)/2019-EMR-I).